\documentclass[12pt]{iopart}

\usepackage{iopams}  

\expandafter\let\csname equation*\endcsname\relax
\expandafter\let\csname endequation*\endcsname\relax

\usepackage{graphicx}
\graphicspath{{./FIGS/}}
\usepackage{dcolumn}
\usepackage{bm}

\usepackage{pgfplots}
\usepackage{arydshln}
\usepackage{mathtools}
\usepackage[utf8]{inputenc} 

\begin{document}

\title{Graph-based analysis of nonreciprocity in coupled-mode systems}
\author{Leonardo Ranzani and Jos\'{e} Aumentado}
\address{National Institute of Standards and Technology, Boulder, CO 80305, USA}
\ead{\mailto{leonardo.ranzani@colorado.edu},\mailto{jose.aumentado@nist.gov}}

\date{\today}

\begin{abstract}
In this work we derive the general conditions for obtaining nonreciprocity in multi-mode parametrically-coupled systems. The results can be applied to a broad variety of optical, microwave, and hybrid systems including recent electro- and opto-mechanical devices. In deriving these results, we use a graph-based methodology to derive the scattering matrix. This approach naturally expresses the terms in the scattering coefficients as separate graphs corresponding to distinct coupling paths between modes such that it is evident that nonreciprocity arises as a consequence of multi-path interference and dissipation in key ancillary modes. These concepts facilitate the construction of new devices in which several other characteristics might also be simultaneously optimized. As an example, we synthesize a novel three-mode unilateral amplifier design by use of graphs. Finally, we analyze the isolation generated in a common parametric multi-mode system, the DC-SQUID.

\end{abstract}



\maketitle


\section{Introduction} \label{section:introduction}
Reciprocity is the symmetry of a physical system with respect to the exchange of a source and detector. For example, in systems which demonstrate nonreciprocal behavior the transmission loss and/or phase delay depends on the direction of propagation~\cite{fan2012comment}. In general, reciprocal symmetry can be violated in multi-mode devices and the most well-known examples are ferrite-based circulators and isolators~\cite{fay1965operation,Pozar2005}. Such devices are useful in several practical contexts: circulators for instance are commonly employed in both optical~\cite{agrawal2012fiber} and cryogenic microwave systems~\cite{clarke2008superconducting,wallraff2004strong,kamal2011noiseless} to reduce reflections and, more importantly, to protect from amplifier backaction on the device-under-test. Likewise, in superconducting quantum information, circulators are often used to direct the amplified reflected signal from lumped-element superconducting parametric amplifiers.

Since the most common schemes for producing nonreciprocity involve the use of high magnetic fields, they are often incompatible with the goal of integrating several measurement circuits into smaller and smaller physical volumes while also limiting the magnetic flux near sensitive superconducting circuits~\cite{kamal2011noiseless}. In the superconducting quantum information and microwave engineering fields, this has motivated a push to understand how nonreciprocity can be generated using alternative methods ~\cite{kamal2011noiseless,fang2013experimental,lira2012electrically,abdo2013,Koch2010} and this general pursuit has been paralleled by similar efforts in optics~\cite{gallo2001all,
longhi2013effective,fang2012realizing,wang2013optical, poulton2012design,sounas2013giant}. Although these different ideas are embodied in different physical implementations, they all share a common mathematical description based in coupled-mode theory. In this paper, we develop a scheme for determining the scattering matrix of an arbitrary coupled-mode system in terms of \emph{directed graphs}. In doing so, we show that nonreciprocity is a consequence of asymmetric interference between different connecting paths in a graph, as well as judiciously-placed dissipation in ancillary modes. We emphasize that while a direct solution of the coupled-mode equations can always be found by means of standard mathematical techniques, the reverse problem of \emph{synthesizing} a multi-mode system with desired properties (gain, impedance match, directionality) can be much more difficult. With the approach we describe in this work, the synthesis of a multi-mode device is translated into the process of building a directed graph whose edges are subject to specific conditions. Since this approach is agnostic to physical implementation, it may benefit similar efforts based on optical, mechanical, and hybrid systems. As many of these systems are parametrically coupled systems, we frame this work in the context of frequency conversion and amplification processes.

This paper is organized as follows--- in Section~\ref{sections:reciprocity} we introduce normalization and matrix conventions for extending the coupled-mode equations of motion to several modes. This facilitates the definition of reciprocity in parametrically coupled systems with three or more modes. In Section~\ref{section:graphrepresentation} we visualize the matrix representation of the equations of motion using directed graphs and show how the scattering matrix can be computed by use of subgraphs which connect the input/output modes. We then use this picture to discuss the mechanism for nonreciprocity in general parametrically-coupled systems. In Sections~\ref{section:threemodeamps} and \ref{section:directionality} we show how directed graphs can be used as both a synthesis and analysis tool. As examples, we synthesize a novel 3-mode parametric amplifier in which gain, directionality, and impedance match are present (Section~\ref{section:threemodeamps}). Finally, we analyze the isolation properties of a complicated system, the DC-SQUID, by casting it as a 10-mode scattering problem (Section~\ref{section:directionality}) based on earlier work by Kamal~\emph{et al.}~\cite{kamal2012gain}. We conclude the paper with a discussion of how this graphical approach might be effectively applied in future work.

\section{Reciprocity in coupled-mode systems} \label{sections:reciprocity}

Resonant and parametrically coupled modes are typically described by the Langevin equations of motion~\cite{louisell1960coupled}. Since in this work we are primarily focused on the \emph{topology} of the coupling in a given system, we utilize a normalized matrix form of these equations and build a graph representation of multimode coupling dynamics. This approach has an additional advantage in that it allows one to deal with systems with several modes in different physical oscillator incarnations as one may encounter in hybrid electro- and opto-mechanical systems. While the transition from the time-domain equations of motion to their matrix representation is given in \textit{Appendix A}, in this section we present only the matrix variables that are necessary for building the graphs.

We begin by considering a set of $N_r$ coupled resonators, described by a set of $N_m$ internal mode amplitudes. In general, we allow the number of modes to exceed the number of resonators $N_m \geq N_r$ such that several modes may reside within a single resonator. This system can be a set of parametrically or resonantly coupled oscillators, but we are not going to assume a particular physical implementation for now, in order to derive results that are as general as possible. We will call $\omega_j$ and $\gamma_j$ the natural oscillation frequency and total dissipation rate for the resonator in which mode $j$ resides. The complex coupling rate between modes $j$ and $k$ will be denoted by $g_{jk}$. In order to describe both mode frequency conversion and amplification, we divide the modes into two subsets of $p$ internal mode amplitudes $b_{1 \dots p}$ and $q=N_m-p$ internal mode amplitudes $b^{\dagger}_{p+1 \dots N_m}$ in the frequency basis, with each set of modes corresponding to the driven response at frequencies $\omega_{1\dots p}^s$ and $-\omega_{p+1:N_m}^s$. We assume that $g_{jk}=g^*_{kj}$ in the case of frequency conversion between modes $j$ and $k$ (for $1 \leq j,k \leq p$ or $p+1 \leq j,k \leq N_m$) and $g_{jk}=-g^*_{kj}$ in the case of parametric amplification (for $1 \leq j \leq p$ and $p+1 \leq k \leq N_m$, or $1 \leq k \leq p$ and $p+1 \leq j \leq N_m$). For example, the coupling matrix for a 2-mode frequency converter or resonantly coupled oscillator system corresponds to the case $p=2, q=0$,  while a conventional parametric amplifier couples two modes, one at a positive frequency ($p = 1$) and the other at a negative frequency ($q = 1$)~\cite{yurkeSU2}. With this general prescription for defining our mode basis, we can perform an input/output analysis of any system of coupled resonators and calculate the corresponding scattering matrix connecting the vector of input fields $b^{in}$ to the vector of output fields $b^{out}$ (see \textit{Appendix A}). The scattering matrix can in general be expressed as 

\begin{equation}
\label{eq:scattering}
S=i\frac{1}{\gamma_M} KM^{-1}K -\mathbb{I}, 
\end{equation} 
where $M$ is a $N_m \times N_m$ matrix given by

\begin{equation}
\label{eq:M_matrix}
M=
\left[
\begin{array}{cccccc}
\Delta_1       & \cdots       & \beta_{1,p}    & \beta_{1,p+1} & \cdots & \beta_{1,N_m} \\
\vdots         & \ddots       & \vdots         & \vdots        &        & \vdots \\
\beta_{1,p}^*  & \cdots       & \Delta_p       & \beta_{p,p+1} & \cdots & \beta_{p,N_m} \\ 
-\beta_{1,p+1}^* & \cdots     & -\beta_{p,p+1}^*   & -\Delta^*_{p+1}      & \cdots    & -\beta_{p+1,N_m}^* \\
\vdots         &              & \vdots             & \vdots               & \ddots    & \vdots \\
-\beta_{1,N_m}^* & \cdots       & -\beta_{p,N_m}^* & -\beta_{p+1,N_m}     & \cdots    & -\Delta_{N_m}^*
\end{array}
\right]
\end{equation} 
and

\begin{align}
\Delta_j &= \frac{\gamma_j}{2\gamma_M}\left (i + \frac{2(\omega^s_j-\omega_j)}{\gamma_j}\right), \label{M_elements1}\\
\beta_{jk} &= \frac{g_{jk}}{2\gamma_M}, \label{M_elements2}
\end{align} 
where we have defined an overall normalization prefactor

\begin{equation}
\gamma_M = {\sqrt[N_m]{\prod_{\ell=1}^{N_m}{\gamma_\ell}}},
\end{equation}
and environmental coupling matrix,

\begin{equation}
K = 
\left[
 \begin{array}{ccc}
\sqrt{\gamma_1^{ext}}  &   & \text{\large0}\\
  &   \ddots &  \\
 \text{\large0}& &  \sqrt{\gamma_{N_m}^{ext}}\\
 \end{array}
\right].
\label{eq:kmatrix}
\end{equation}

In general $\gamma_j \geq \gamma_j^{ext}$, with the equal sign if the internal dissipation rate of mode $j$ is zero. The diagonal elements of $M$ are complex normalized detunings between the driven response frequencies and the natural resonator frequencies and include the mode dissipation rates, while the off-diagonal elements are normalized coupling coefficients. In principal, the normalizations above are unnecessary, but they allow one to cast the Langevin equations of motion (see \textit{Appendix A}) in a simple matrix form (as $M$, the ``Langevin matrix''), emphasizing the underlying structure of the connections between modes. We show below that this structure, as revealed in the graph representation of $M$, allows one to draw immediate conclusions about the (non)reciprocity in multimode coupled systems.

\paragraph*{A formal definition of reciprocity.}
In order to proceed, we must formally define reciprocity within the context of several modes, possibly oscillating at different frequencies (as found in parametrically coupled systems). The scattering matrix (Equation~\ref{eq:scattering}) describes a fully reciprocal system (transmission between \emph{any} two modes is reciprocal) if it obeys the general constraint~\cite{deak2012reciprocity,Leung2010},
 
\begin{equation}
\label{eq:reciprocity}
S^T=U_{\phi} S U^\dagger _{\phi}.
\end{equation}

Here the matrix $U_{\phi}$ corresponds to a set of phase shifts in the mode basis

\begin{equation}
\label{phase shift}
U_{\phi} = 
\left[
 \begin{array}{ccc}
e^{i\phi_1}  &   & \text{\large0}\\
  &   \ddots &  \\
 \text{\large0}& &  e^{i\phi_{N_m}}\\
 \end{array}
\right],
\end{equation}
where we can assume without loss of generality that $\sum \limits_{\ell=1}^{N} {\phi_\ell}=0$. The matrix $U_{\phi}$ therefore corresponds to an arbitrary redefinition of the mode phases. This is similar to the concept of a common mode phase shift, but generalized to several modes. 

A system that violates condition~(\ref{eq:reciprocity}) is nonreciprocal and is characterized by an asymmetric phase and/or amplitude transmission coefficient between at least one pair of modes. We note that Equation~\ref{eq:reciprocity} differs from the definition commonly found in textbooks\cite{Pozar2005}, $S^T = S$. In fact, Equation~\ref{eq:reciprocity} is valid for more general multimode parametric systems in which the mode frequencies can differ. As such, it is a gauge-invariant definition of nonreciprocity~\cite{deak2012reciprocity,Leung2010}. Since the environmental coupling matrix $K$ in Equation~\ref{eq:kmatrix} is diagonal, it follows from Equation~\ref{eq:scattering} that reciprocity is guaranteed if and only if the Langevin matrix, $M$, obeys the same similarity transformation as in Equation~\ref{eq:reciprocity},

\begin{equation}
\label{eq:reciprocityH}
M^T=U_{\phi} M U^\dagger _{\phi}, 
\end{equation}
for some set of phase shifts $\{\phi_\ell\}$. Above, we noted that all resonant and parametric systems considered here are constrained to couplings which are related by pairwise (anti)-conjugation, \emph{i.e.}, $\beta_{jk}=\pm \beta^*_{kj}$. Placing this constraint in Equation~\ref{eq:reciprocityH} we obtain

\begin{equation}
\label{eq:reciprocityH2}
\phi_j-\phi_k=-2 \angle{\beta_{jk}} + 2n_{jk} \pi, 
\end{equation}
where $\angle{\beta_{jk}}$ is the phase of the coupling coefficient $\beta_{jk}$ between modes $j$ and $k$ and $n_{jk}$ is an integer. The system of Equations~\ref{eq:reciprocityH2} has at most a set of $N_m(N_m-1)/2$ independent relations (if all off-diagonal elements of $M$ are nonzero) in $N_m-1$ unknown phases $\{\phi_\ell\}$. The system is therefore overdetermined and the reciprocity condition (Eqn.~\ref{eq:reciprocity}) is, for parametrically coupled systems, satisfied only for specific choices of coupling phases.

\begin{figure}
\begin{center}
{\includegraphics[width=16 cm,trim=0 0 0 0cm]{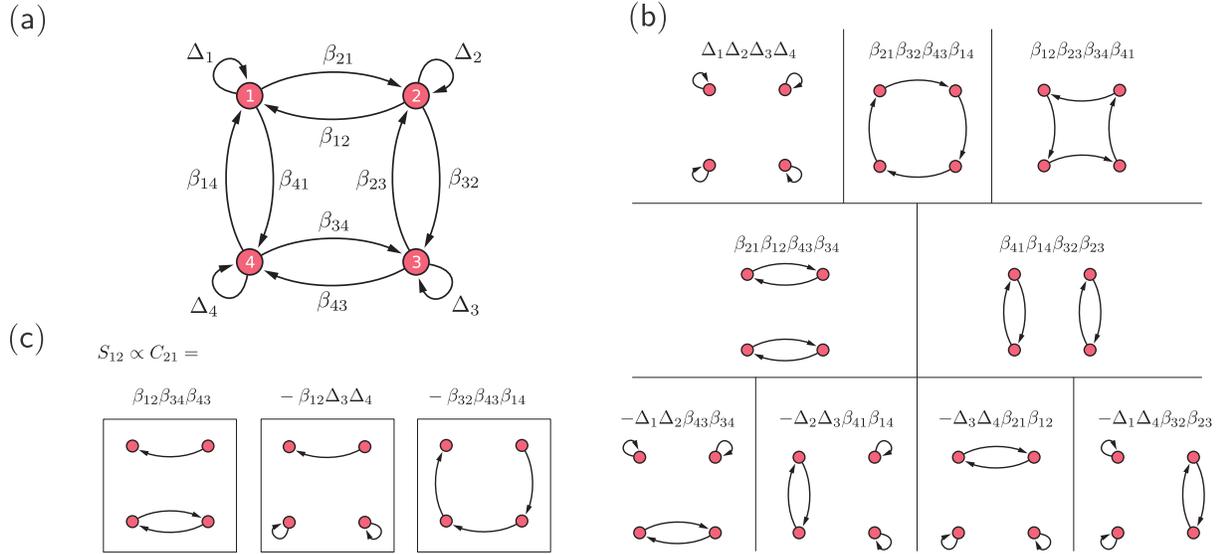}}
    \caption{(a) Graph representation of a 4-mode Langevin matrix $M$ in~(\ref{eq:M_matrix}). $M$ is the adjacency matrix of this graph, \emph{i.e.}, every element $M_{jk}$ in $M$ is the weight of an oriented edge from vertex $k$ to vertex $j$ in the graph. 
	(b) Calculation of the determinant $|M|$ and (c) the cofactor $C_{jk}$ by use of permutation subdigraphs. The cofactor $C_{jk}$ is proportional to the scattering coefficient $S_{kj}$ (\emph{see text}).}
\label{Fig:determinant_calculation}
\end{center}
\end{figure}

\section{Graph representation of the Langevin matrix} \label{section:graphrepresentation}

The conditions for Equations~\ref{eq:reciprocityH2} to have a unique solution, and therefore for the system to be reciprocal, can be visualized by drawing a \emph{directed graph} representing the matrix $M$ (see Figure~\ref{Fig:determinant_calculation}). The graph has $N_m$ vertices and we associate a weighted directed branch (``edge'') between vertices $j$ and $k$ to every element $\beta_{jk}$ in the matrix $M$. The diagonal elements $\Delta_j$ and $-\Delta^*_k$ of $M$ are the weights of self-loops in the graph. In other words, all of the coupling terms in $M$ are shown as directed branches, and all mode detuning and dissipation is represented by self-loops in the graph.

Within this representation, the conditions for a unique solution for the entire system can be obtained by summing Equations~\ref{eq:reciprocityH2} along every closed loop $L$, yielding a simple result,

\begin{equation}
\label{eq:reciprocityH3}
\sum \limits_{\beta_{jk}\in L} \angle{\beta_{jk}}=k_L \pi, 
\end{equation}

where $k_L$ is an integer. Equation~\ref{eq:reciprocityH3}, reminiscent of Kirchoff's first law in electrical circuits, is analogous to the reciprocity condition in Jaynes-Cummings lattices described in~\cite{Koch2010,Koch2011}. We stress, however that Equation~\ref{eq:reciprocityH3} is applicable to any linear coupled-mode system bearing arbitrarily complex connections and is not confined to the nearest-neighbor, planar graph couplings shown here. Moreover, in the case of parametrically coupled devices in particular, the vertices of our graphs represent modes that may physically correspond to energy that is differentiated by both space and frequency.  As such, the graphs are a very general representation of a coupled mode system and can describe coupling of energy at several different frequencies as well as physical ports.

In order for a system to be \emph{nonreciprocal}, the condition \ref{eq:reciprocityH3} must be violated. This, however, only expresses a minimal requirement for phase nonreciprocity. We must further define a condition for which a system will also demonstrate amplitude nonreciprocity. Specifically, we may ask which conditions allow the amplitude isolation ratio $|S_{jk}/S_{kj}|$ to be different than 1. From (\ref{eq:scattering}) it follows that the scattering coefficient $S_{jk}$ scales with the corresponding element of the inverse Langevin matrix, $M^{-1}_{jk}$. The problem of calculating the scattering coefficients is therefore reduced to the problem of inverting the Langevin matrix which, as we show below, can also be conveniently cast in terms of directed graphs. 

The inverse Langevin matrix can be expressed in terms of the transpose of its cofactor matrix~\cite{anthony2012linear}

\begin{equation}
M^{-1} = \frac{C^T}{|M|}.
\label{eq:inverseMcofactor}
\end{equation} 

The scattering coefficient is then given by

\begin{equation}
S_{jk}=i\frac{\sqrt{\gamma^{ext}_j\gamma^{ext}_k}}{\gamma_M}\frac{C_{kj}}{|M|}-\delta_{jk},
\label{eq:scattering2}
\end{equation} 

where $\delta_{jk}$ is the Kronecker delta function. The elements of the cofactor matrix $C_{kj}$ are equal to the determinant of the matrix obtained from $M$ after removing the $k$-th row and the $j$-th column (\emph{i.e.}, the $kj$ minor), multiplied by $(-1)^{j+k}$~\cite{anthony2012linear}. 

The determinant of $M$ and the cofactor $C_{kj}$ can be easily calculated from the graph in Figure~\ref{Fig:determinant_calculation}(a) by the following procedure (\emph{see} Refs~\cite{greenman1976graphs,brualdi2008combinatorial}). Given a graph, we call a subgraph $G$ a ``permutation subdigraph'' if every vertex in $G$ has one and only one outgoing edge and one and only one incoming edge~\cite{brualdi2008combinatorial}. The total weight $w(G)$ of a subgraph $G$ is the product of the weights of all of its edges times  $(-1)^{c+N_m}$, where $c$ is the number of cycles in $G$ (by cycle we mean a sequence of distinct connected vertices, where the first and last vertex coincide). The determinant of the matrix $M$ is the sum of weights of every possible permutation subdigraph of the graph associated to $M$~\cite{greenman1976graphs,brualdi2008combinatorial}, as shown in Figure~\ref{Fig:determinant_calculation}(b).

The cofactor matrix $C_{kj}$ is the coefficient of the $M_{kj}$ element of the matrix $M$ in the Laplace expansion of the determinant $|M|$. Therefore, in order to calculate $C_{kj}$, we need to compute only the permutation subdigraphs in the expansion of $|M|$ that contain the $kj$ edge, with the $kj$ edge removed. In other words, for every path $p$ FROM vertex $k$ TO vertex $j$, we consider the collection of all the permutation subdigraphs $G^p_r$ of the graph of $M$ containing $p$, with the $kj$ edge removed as in Figure~\ref{Fig:determinant_calculation}(c). The cofactor $C_{kj}$ is then given by

\begin{equation}
\label{minor_eq}
C_{kj}=-\sum \limits_{p,r}{w(G^p_r)}, 
\end{equation} 
where $p$ varies over all the paths connecting vertex $j$ to vertex $k$ and $r$ varies over all the subdigraphs containing the path $p$. The minus sign is due to the fact that, by removing the $kj$ edge, we reduced the number of cycles in the subdigraph by 1. An example of this procedure is outlined in Figure~\ref{Fig:determinant_calculation}(b). The graphs $\{G^p_r\}$ in the expression for the cofactor elements have a simple physical interpretation: they represent the possible scattering mechanisms that connect modes $j$ and $k$.

Since, from (\ref{eq:scattering2}), the ratio between the scattering coefficients $S_{jk}/S_{kj}$ is equal to the ratio between the corresponding cofactors $C_{kj}/C_{jk}$, the problem of analyzing isolation in a given multi-mode system can be reduced to evaluating the differences between the specific graphs that connect $j$ to $k$ and \textit{vice versa}.

\begin{figure}
\begin{center}
{\includegraphics[width=16 cm,trim=0 0 0 0cm]{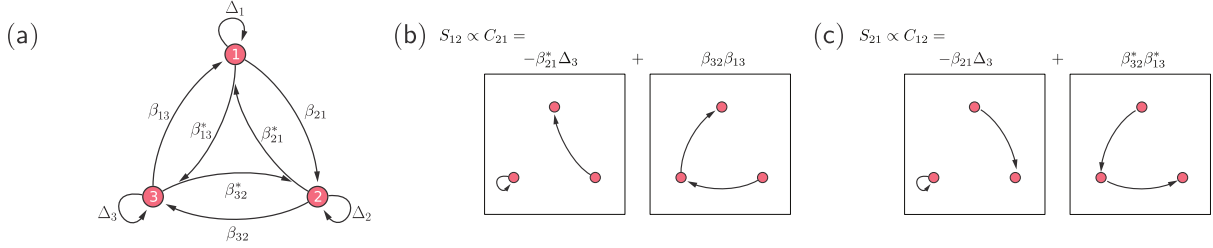}}
    \caption{(a) Graph representation of a 3-mode system and calculation of the scattering coefficients $S_{12}$ (b) and $S_{21}$ (c). The element $C_{12}$ of the cofactor matrix is calculated by summing the weights of the permutation subdigraphs containing a path from vertex $1$ to vertex $2$, with the edge from vertex $2$ to vertex $1$ removed.}
\label{Fig:determinant_3mode}
\end{center}
\end{figure}
 
As a more concrete example, we apply this procedure to a simple three-mode system in which all modes are coupled via frequency conversion ($\beta_{jk} = \beta_{kj}^*$) and are therefore described by the following matrix (see Figure~\ref{Fig:determinant_3mode}),

\begin{equation}
\label{3mode_eq}
M=
\left[\begin{array}{ccc}
\Delta_1 & \beta_{12} & \beta_{13} \\
\beta^*_{12} & \Delta_2 & \beta_{23} \\
\beta^*_{13} & \beta^*_{23} & \Delta_3 
\end{array} \right],
\end{equation} 

where $\Delta_j$ and $\beta_{jk}$ are normalized detunings and couplings as before. The isolation between ports $1$ and $2$ can be calculated by use of the graphs representing the cofactors (see Figure~\ref{Fig:determinant_3mode}(b) and (c)) yielding the expression,

\begin{equation}
\label{eq:iso3mode}
I_{12}=\frac{S_{12}}{S_{21}}=\frac{C_{21}}{C_{12}}=\frac{-\beta_{21}^*\Delta_3 + \beta_{32}\beta_{13}}{-\beta_{21}\Delta_3 + \beta_{32}^*\beta_{13}^*}=\frac{-g_{21}^*(\omega_3^S - \omega_3 + i\gamma_3/2) + g_{32}g_{13}}{-g_{21}(\omega_3^S - \omega_3 + i\gamma_3/2) + g_{32}^*g_{13}^*},
\end{equation}
where, in the last line, we have removed all normalizations.

From (\ref{eq:iso3mode}) we observe that if the total dissipation of mode ``3'', $\gamma_3=0$, then $S_{12}=S^*_{21}$ and there is no isolation between modes 1 and 2. In this case mode 3 serves as an internal lossless mode and the system behaves as a simple reciprocal two-mode system. Here, dissipation plays a crucial role in violating reciprocity, as it breaks the symmetry between scattering elements which are otherwise related by complex conjugation. Assuming finite dissipation in mode 3, we find the condition for maximum isolation $I_{12} = 0$,

\begin{equation}
\label{eq:maxiso3}
\frac{g_{13}g_{32}}{g_{21}^*}=\omega_3^s-\omega_3+i\gamma_3/2.
\end{equation}

In other words, to achieve maximum isolation, the sum of the phases of the couplings along the loop connecting the three modes, $\phi_{13} + \phi_{21} + \phi_{32}$, must be equal to the detuning angle, $\tan^{-1}[\gamma_3/(2(\omega_3^s-\omega_3))]$. At zero detuning in mode 3, $\omega_3^s =\omega_3$, this loop phase must be $\pi/2$ for perfect isolation. Equivalently, one can interpret this isolation condition, Eqn.~\ref{eq:maxiso3}, as interference between the two possible coupling paths connecting modes 1 to 2 in the graph (Figure~\ref{Fig:determinant_3mode}). Although we focus on the isolation here, condition \ref{eq:maxiso3} also yields unit transmission in the forward direction, $|S_{21}| = 1$, for $|g_{12}|\gamma_3=1$. In fact, one can continue to build a 3-mode circulator by finding similar constraints for the 2$\leftrightarrow$3 and 1$\leftrightarrow$3 coupling.

\section{Three-mode directional amplifiers} \label{section:threemodeamps}

\begin{figure}
\begin{center}
{\includegraphics[width=16 cm,trim=0 0 0 0cm]{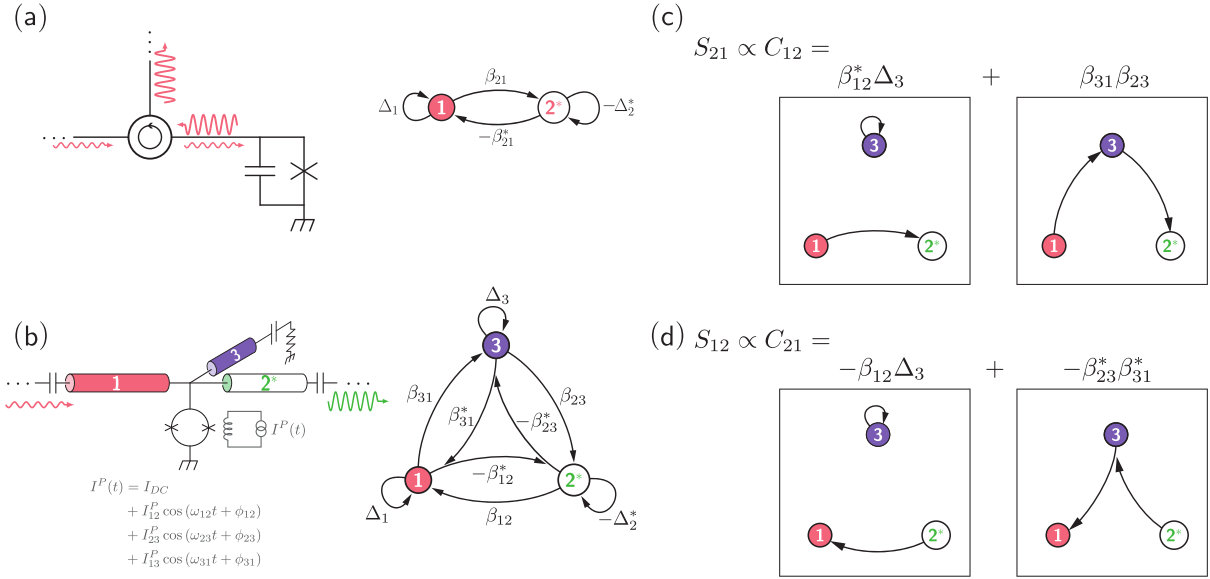}}
    \caption{(a) A traditional Josephson parametric amplifier, unidirectionality is achieved with an auxiliary circulator.  (b) Three-mode Delta amplifier discussed in this work, which consists of one flux-biased DC-SQUID connected to 3 resonators at frequencies $\omega_{1,2,3}$ and its graph representation. The SQUID is equivalent to a nonlinear inductor modulated by the magnetic flux through the SQUID loop. One pump at frequency $\omega_{p1}=\omega_2+\omega_1$ is used to obtain parametric amplification between mode 1 at frequency $\omega_1$ and mode 2 at frequency $\omega_2$. The second pump at frequency $\omega_{p2}=\omega_3-\omega_1$ generates frequency conversion between mode 1 and mode 3 at frequency $\omega_3$. A third pump at frequency $\omega_{p3}=\omega_{p1}+\omega_{p2}$ closes the loop and generates parametric amplification between modes 2 and 3. In this configuration non-reciprocal trans-gain can be obtained between modes 1 and 2. (c) and (d) subgraphs used to calculate the forward and reverse gain between modes 1 and 2.}
\label{Fig:3mode_amp}
\end{center}
\end{figure}

Traditional single-stage lumped-element parametric amplifiers, such as the one shown in Figure~\ref{Fig:3mode_amp}(a), operate in reflection-mode such that a circulator is needed in order to achieve unidirectionality. In~\cite{abdo2013josephson} a single-stage amplifier with two parametric pumps at the same frequency was introduced, but in addition to forward gain it also had unity reverse gain. The graph methodology introduced here allows us to revisit the problem of gain and isolation from another perspective. In particular, our discussion above identified the importance of introducing a multi-edge loop to obtain nonreciprocity. Here we analyze a closed-loop amplifier which we term the ``Delta'' amplifier, shown schematically in Figure~\ref{Fig:3mode_amp}(b) both as a device concept and as its corresponding graph. This amplifier can provide forward gain and zero reverse gain as discussed below.

As drawn, the Delta amplifier requires three independent pumps to generate three pairs of coupling edges $\{\beta_{jk}\}$. Two of these pairs create amplification (\emph{i.e.,} $\beta_{23} = -\beta_{32}^*$, $\beta_{12} = -\beta_{21}^*$), while the third pair couples through frequency conversion ($\beta_{31} = \beta_{13}^*$). One concept for how this might be implemented is shown schematically in Figure~\ref{Fig:3mode_amp}(b) as three resonators that share a common current anti-node in which a flux-driven SQUID is operated as a parametrically modulated inductor. In principle, one might drive two of these couplings using the same pump frequency \footnote{We recently learned of a similar three-mode coupled system driven with a ``biharmonic'' pump, outlined by Kamal \emph{et al.}~\cite{kamal2014asymmetric}.}\cite{kamal2014asymmetric}, but for clarity we discuss only the case in which each of the three couplings is driven independently, which has the advantage of increased flexibility due to the increased number of degrees of freedom. Regardless of particular implementation, we can recognize the presence of a closed loop connecting all three modes, as well as the integration of amplification as a coupling process. From this bare description we recognize that with three independent pumps we are constrained to optimizing only three of the nine possible scattering elements. In laboratory use, we can choose to optimize for amplification in the ``forward'' direction ($|S_{21}|>1$), isolation in the ``reverse'' direction ($S_{12}\sim 0$), and low input reflection coefficient (input match, \emph{i.e.}, $S_{11}\sim 0$) to prevent unwanted reflections of signals propagating away from a device-under-test.

We begin by writing down the expressions for the forward and reverse transmission from Equation~\ref{eq:scattering2},

\begin{align}
\label{delta_gain}
S_{21} &= i\frac{\sqrt{\gamma^{ext}_1\gamma^{ext}_2}}{\sqrt[3]{\gamma_1\gamma_2\gamma_3}}\frac{\beta^*_{12}\Delta_3+\beta_{31}\beta_{23}}{|M|} \\
S_{12} &= i\frac{\sqrt{\gamma^{ext}_1\gamma^{ext}_2}}{\sqrt[3]{\gamma_1\gamma_2\gamma_3}}\frac{-\beta_{12}\Delta_3 - \beta^*_{23}\beta_{31}^*}{|M|},
\end{align} 
where $|M|$ is the determinant of the coupling matrix represented in the graph shown in Figure~\ref{Fig:3mode_amp}(b). Assuming ideal isolation, $S_{12} = 0$, we find an expression similar to Equation~\ref{eq:maxiso3}, 

\begin{equation}
\label{isodelta}
\frac{\beta_{23}^*\beta^*_{31}}{\beta_{12}}=-\Delta_3,
\end{equation}
which provides an overall constraint on the coupling amplitudes and phases as a function of the complex detuning of mode 3. Under this primary constraint, we then set the input reflection coefficient to zero,

\begin{equation}
\label{deltas11}
S_{11}=i\frac{\gamma^{ext}_1}{\sqrt[3]{\gamma_1\gamma_2\gamma_3}}\frac{|\beta_{23}|^2-\Delta^*_2 \Delta_3}{|M|}-1=0,
\end{equation}
giving a constraint on the determinant,

\begin{equation}
\label{delta_det}
|M|=i\frac{\gamma^{ext}_1}{\sqrt[3]{\gamma_1\gamma_2\gamma_3}}(|\beta_{23}|^2-\Delta^*_2 \Delta_3).
\end{equation}

Taken together, these constraints yield an expression for the forward transmission gain, 

\begin{equation}
\label{deltagain}
\sqrt{G}=|S_{21}|=\sqrt{\frac{\gamma^{ext}_2}{\gamma^{ext}_1}}\frac{2|\beta_{12}\mathrm{Im}(\Delta_3)|}{\bigl||\beta_{23}|^2-\Delta^*_2\Delta_3\bigr|}.
\end{equation}

If external coupling dominates, $\{\gamma_k\}=\{\gamma^{ext}_k\}$, in the limit of small detunings $(\omega^s_k-\omega_k) \ll \gamma_k$ the gain reduces to

\begin{equation}
\label{deltagain_simplified}
\sqrt{G} \approx \sqrt{\frac{\gamma_2}{\gamma_1}}\frac{2|g_{12}|\gamma_3}{|g_{23}|^2-\gamma_2\gamma_3}.
\end{equation}

\begin{figure*}
\begin{center}
{\includegraphics[width=16 cm,trim=0 0 0 0cm]{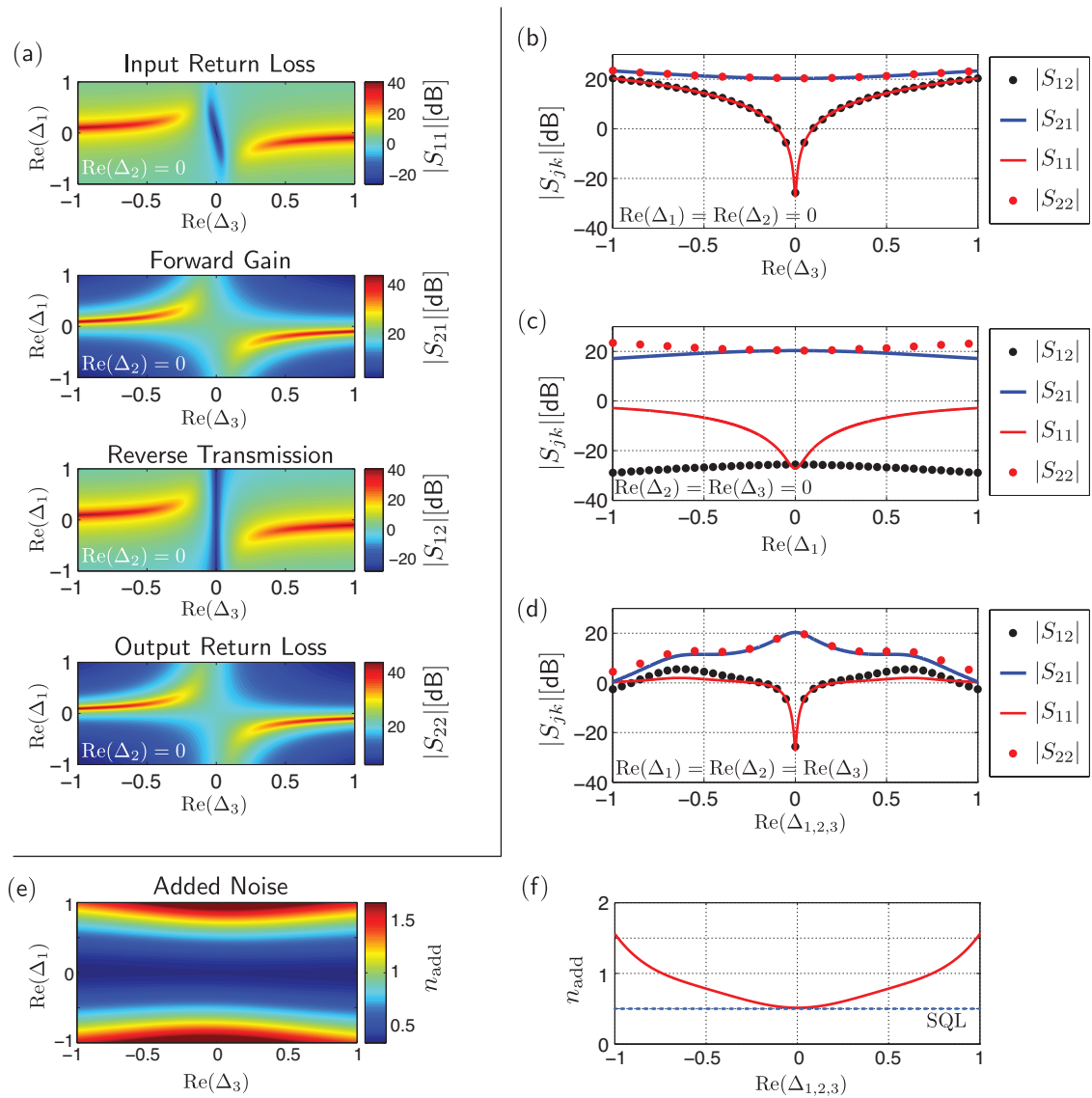}}
    \caption{(a) Simulated input return loss, forward and reverse gain of the Delta amplifier in Figure~\ref{Fig:3mode_amp} as a function of normalized mode 1 and 3 detunings for $G$=20~dB and same dissipation rate for the 3 modes $\gamma_k=\gamma^{ext}_k$. The couplings were set according to Equations~(\ref{isodelta}),(\ref{delta_det}.) and~(\ref{deltagain_simplified}). In (b) and (c) the scattering parameters are plotted as a function of modes 1 and 3 detunings respectively. (d) is obtained if the pump frequencies are kept constant and only the input signal frequency is swept. (e)\&(f) Amplifier added noise appearing at port 2, referred to port 1. (f) is an equal-detuning line-cut (similar to (d)) showing added noise (red curve) reaching the standard quantum limit (SQL, blue dashed line) for this gain.}
\label{Fig:Delta_sym}
\end{center}
\end{figure*}

Note that this form gives the gain at a fixed frequency where the input match is perfect (Equation~\ref{deltas11}). By substituting Equations~(\ref{delta_gain}) and~(\ref{isodelta}) in Equation~(\ref{deltas11}) at resonance we can determine the three coupling rates. In the limit of high gain, $|g_{jk}|\approx \sqrt{\gamma_j\gamma_k}$, with the exact value depending on the gain $G$.

We have calculated the resulting $S$-parameters for the Delta amplifier in Figure~\ref{Fig:Delta_sym} under various constraints on the pump detunings for the case of uniform dissipation rates $\gamma_k=\gamma$. In practice, one might fix the pump frequencies, locking the three detunings together, and measure the $S$-parameters shown in Figure~\ref{Fig:Delta_sym}(d). For this example calculation, we selected values for the pump amplitudes to fix the nominal power gain to $\sim$20~dB while obtaining significant isolation. The device bandwidth is determined according to Equation~(\ref{delta_gain}) that fixes the gain-bandwidth product for specific dissipation rates. We note here that the device bandwidth is comparable to traditional two-mode superconducting lumped parametric amplifiers, while the center frequency could be tuned by changing the frequencies of the pumps. Finally, we find that the calculated output $S_{22}$ shows reflection gain at the output port, and is consistent with the limitations of having only three parameters (the independent pumps) to optimize three of the four scattering parameters between ports 1 and 2. However, this might not be a limitation in practical use where one is more concerned with the backaction of the amplifier on a quantum circuit on the input side (port 1).

\paragraph*{Added noise of the three-mode amplifier.} We computed the equivalent input added noise $\bar{n}_{add}$ of the Delta amplifier to check that the device can approach the standard quantum limit. We assumed that half a photon of noise is injected into every port and calculated the total output noise spectral density from the scattering parameters. The contribution from ports $2$ and $3$, divided by the device power gain $G$, constitutes the equivalent added input noise, as outlined in~\cite{metelmann2014quantum}. For a symmetric device, in the limit of large gain and at zero detuning, the result is:

\begin{equation}
\bar{n}_{add}(0)=\frac{|S_{22}(0)|^2+|S_{23}(0)|^2}{2G} \approx \frac{|S_{22}(0)|^2}{2G} \approx \frac{1}{2}\left ( 1+\frac{1}{\sqrt{G}} \right )^2,
\end{equation}  

where we can neglect $|S_{23}|$ as a consequence of the isolation condition. The results of the numerical calculations for nonzero detuning are shown in Figure~\ref{Fig:Delta_sym}. The fact that the Delta amplifier is quantum-limited may be surprising at first, as one might expect the third mode to contribute extra noise beyond the quantum limit. However the third mode acts only as an effective dissipative coupling between modes $1$ and $2$~\cite{metelmann2014quantum}. We note that the parametric amplification scheme that is the subject of Ref.~\cite{metelmann2014quantum} constitutes a three mode device in which the added noise was determined to be at the standard quantum limit. In fact, the Delta amplifier is similar to that device except for the addition of an amplification branch which serves to create a closed loop in its graph description, introducing the necessary interference for directionality.

\section{Directionality of DC-SQUID amplifiers} \label{section:directionality}

\begin{figure}
\begin{center}
{\includegraphics[width=12 cm,trim=0 0 0 0cm]{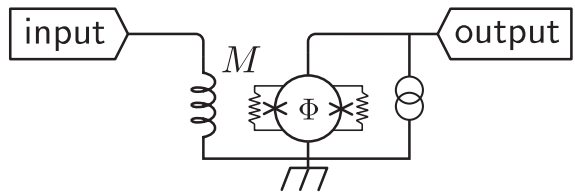}}
    \caption{Schematic of a DC-SQUID amplifier. The SQUID can be modelled as a parametric coupled-mode system. The modes at frequencies $\omega \pm n\omega_J$ are coupled through internally generated pump signals at multiples of the Josephson frequency $\omega_J$. At each frequency there are two doubly degenerate modes, corresponding to the common and differential excitation of the junction phases.  }
\label{Fig:squid}
\end{center}
\end{figure}

While the previous discussion focused on using a graphical approach to synthesize a novel amplifier, it can also be used to identify  aspects of other systems which enable both gain and directionality. In this section we use graphs to analyze a complicated multi-mode system, the DC-SQUID amplifier. DC-SQUIDs are a type of superconducting amplifier with noise temperatures approaching the standard quantum limit~\cite{defeo:092507,Muck2010,spietz2009superconducting}. Since DC-SQUIDs also have inherently low power dissipation (in addition to high gain and directionality~\cite{ranzani2013broadband}), they have been the subject of several applied superconductivity efforts to produce reliable amplifiers for microwave quantum information measurements. In Ref.~\cite{kamal2012gain}, Kamal \emph{et al.} showed that the directionality of the SQUID is a consequence of multiple parametric amplification and frequency conversion processes due to frequency mixing between the input microwave signal and internally-generated parametric pumps at multiples of the Josephson frequency. This approach casts the DC-SQUID as a multi-mode coupling problem and, as such, is also amenable to a graph description. Here we describe the parametric coupling in the SQUID following the description in Ref.~\cite{kamal2012gain} by use of graphs to map the connections between modes. In particular, we will direct our attention to the interference and dissipation that must be present to generate nonreciprocity in the presence of gain.

\begin{figure*}
\begin{center}
{\includegraphics[width=16 cm,trim=0 0 0 0cm]{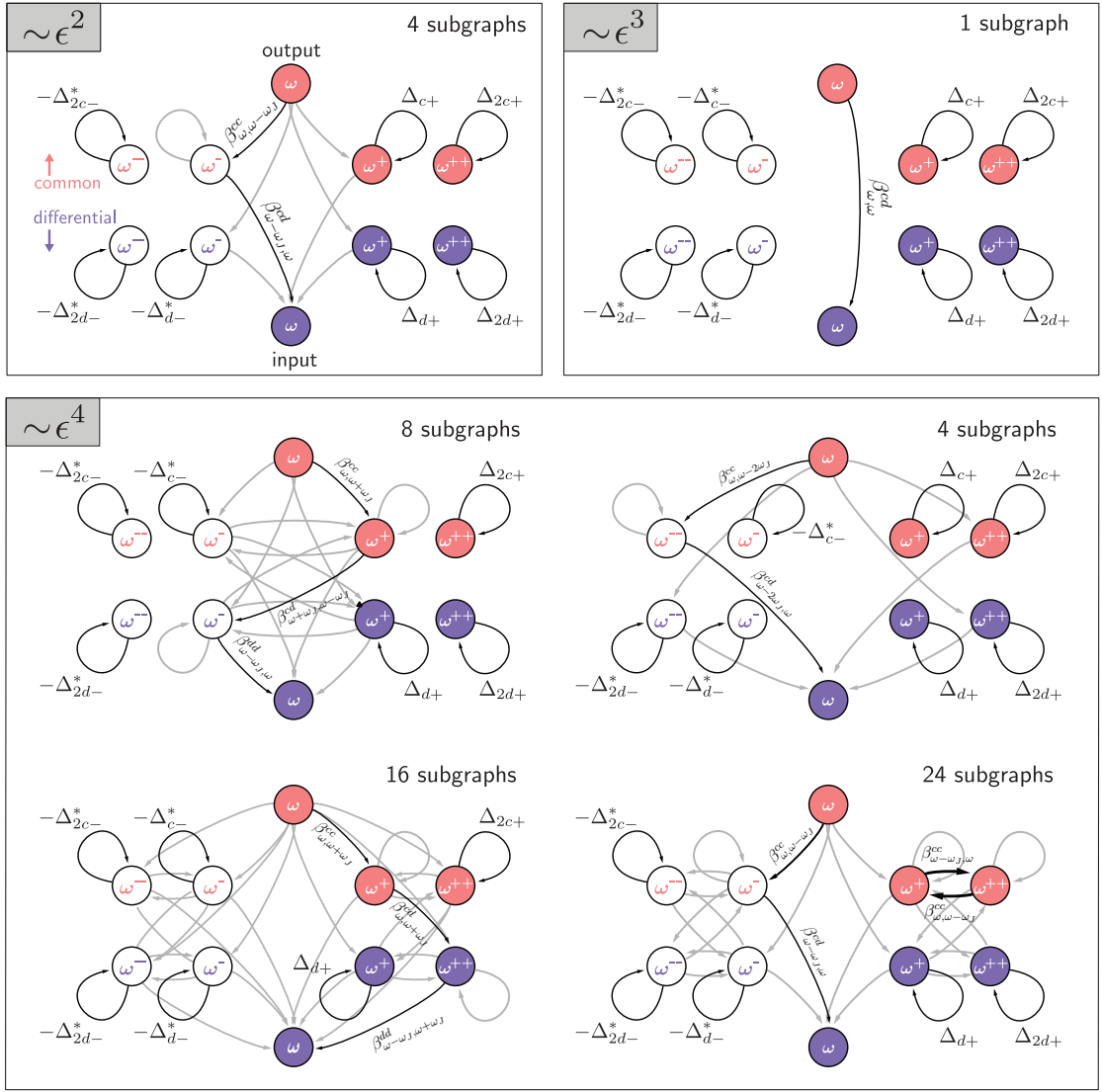}}
    \caption{Leading-order subgraphs describing mode coupling in the SQUID. For simplicity we show only one typical subgraph for each order in $\epsilon$. The other subgraphs are obtained by simple permutation of the edges shown. The common modes are shown in red, while the differential modes are shown in purple.}
\label{Fig:squid_graph_approximate}
\end{center}
\end{figure*}

A DC-SQUID consists of a pair of Josephson junctions inside a superconducting loop as in Figure~\ref{Fig:squid}. Each junction is characterized by a phase difference $\phi_{1,2}$ across its terminals, but it is more convenient to recast the dynamics of the DC-SQUID in terms of the common and differential phase difference $\phi_{C,D}=\phi_1 \pm \phi_2$~\cite{kamal2012gain}. $\phi_{C,D}$ can be expressed as the sum of fast oscillating terms at multiples of the Josephson frequency $\omega_J=2\pi V_{dc} / \Phi_0$, where $\Phi_0$ is the magnetic flux quantum, and slow oscillating terms at the signal frequency $\omega$. The fast oscillating terms act as parametric pumps causing mode conversion between $\omega$ and the mixing product frequencies $\omega \pm k\omega_J$. The lowest-dimensional model that can capture nonreciprocity involves two parametric pumps and a total of 10 modes at 5 frequencies $\omega \pm 2\omega_J,\omega \pm \omega_J, \omega$. Each mode is doubly-degenerate, because for each frequency there are a common and a differential mode excitation. Some scattering subgraphs to describe the SQUID are shown in Figure~\ref{Fig:squid_graph_approximate}, where, for example, $\beta^{cc}_{\omega,\omega-\omega_J}$ indicates the coupling rate between the common modes at frequency $\omega$ and $\omega-\omega_J$. The input and output modes are shown at the very bottom and top respectively. Calculation of the mode coupling rates as a function of the internally generated pump amplitudes is detailed in \textit{Appendix B}.

The number of subgraphs to describe the SQUID is too large for analytical calculations. However if the SQUID bias current is high enough, we can look for an approximate expression for the SQUID isolation for small values of the bias parameter $\epsilon=I_c/I_b$, where $I_c$ is the critical current of the junctions and $I_b$ is the bias current. The pump amplitudes and the coupling coefficients can then be calculated for small $\epsilon$ and we can simplify our calculations by considering only the subgraphs of leading order in $\epsilon$. These subgraphs are shown in Figure~\ref{Fig:squid_graph_approximate}. The order of the graph is given by the product of its edges and therefore depends on the length of the path connecting the input to the output modes as well as the order of the edge weights (coupling rates). The order of the coupling rates increases with the difference between the mode drive frequencies (for example $\beta^{cc}_{\omega,\omega-k\omega_J} \sim \epsilon^k$). Moreover a small ($\sim \epsilon^3$) direct coupling exists between modes at the same frequency, as explained in \textit{Appendix B}. 

\begin{figure}
\begin{center}
{\includegraphics[width=16 cm,trim=0 0 0 0cm]{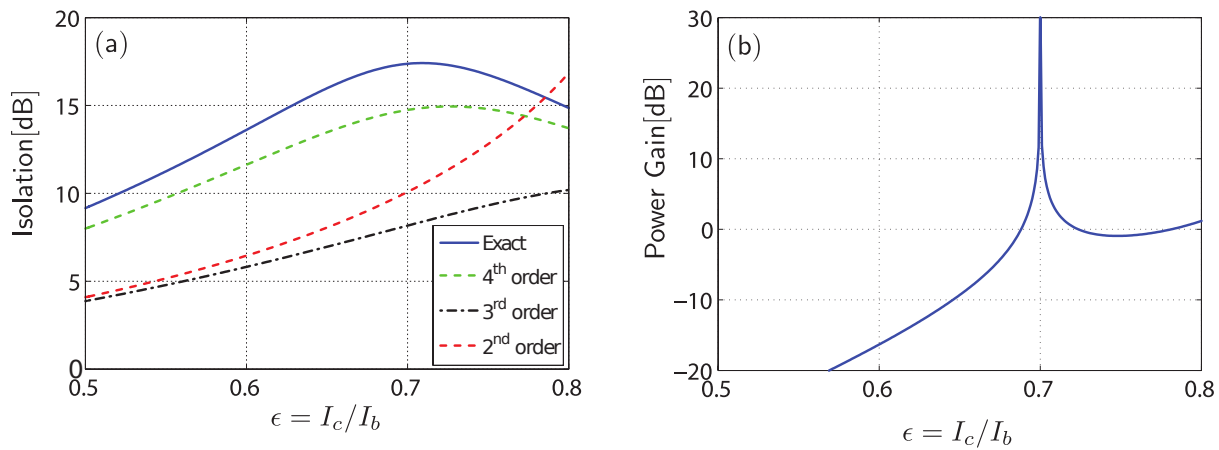}}
    \caption{(a) Comparison between the exact calculation of the SQUID isolation ($|S_{CD}/S_{DC}|$) based on the linearized coupled-mode equations and the approximate solutions including subgraphs up to the $2^{nd}$, $3^{rd}$, and $4^{th}$ order for $\phi_e=\pi/4$, $\beta_l=1$ and $\omega=0.01\omega_J$. (b) Exact power gain.}
\label{Fig:squid_graph_approximate2}
\end{center}
\end{figure}

The dissipation rates for the common and differential modes when the SQUID is 
terminated into ideal infinite transmission lines, are given by

\begin{align}
\label{gammas}
\gamma_C &=\frac{1}{RC}=\frac{\omega_c}{\beta_c},\\
\gamma_D &=\frac{2R}{L}=\frac{2\omega_c}{\pi\beta_L},
\end{align}

with $\beta_L=(2 L I_c) / \Phi_0$, $\beta_c=2\pi I_c R^2 C / \Phi_0$, $\omega_c = 2 \pi I_c R /\Phi_0$, $L$ is the inductance of the SQUID loop, and $R$ and $C$ are the junction shunt resistance and capacitance. Since the SQUID is a nonresonant system, the mode frequencies $\omega_i$ in Equation~(\ref{M_elements1}) are zero. We can approximately compute the SQUID isolation by use of the lowest-order subgraphs shown in Figure~\ref{Fig:squid_graph_approximate}. The first conclusion we can draw, based on our previous discussion, is that non-zero dissipation rates in (\ref{gammas}) are essential to obtain nonreciprocity in the SQUID. A change in the impedance terminations at the various mode frequencies $\omega \pm k\omega_J$ will change the amount of nonreciprocity, \emph{viz.}, modes that are terminated in a short or an open circuit (corresponding to zero dissipation rates) will not contribute to the SQUID isolation. Moreover, it is important that $\gamma_C \neq \gamma_D$. In fact, by taking into account the symmetries between the mode couplings, the weight of the $2^{nd}$ order subgraphs in Figure~\ref{Fig:squid_graph_approximate} is given by

\begin{equation}
\label{lowest_order}
w_{\epsilon^2} \propto 4\omega\operatorname{Re}\left[\beta^{cc}_{\omega+\omega_J,\omega}\beta^{cd}_{\omega,\omega+\omega_J}\right]+4(\gamma_C-\gamma_D)\operatorname{Im}\left[\beta^{cc}_{\omega+\omega_J,\omega}\beta^{cd}_{\omega,\omega+\omega_J}\right],
\end{equation}

apart for a common factor equal to the product of the internal modes normalized detunings. If $\gamma_C=\gamma_D=\gamma$, the second term in Equation~\ref{lowest_order} is zero and the SQUID becomes reciprocal at first order. The same cancellation happens in the $3^{rd}$- order subgraphs in Figure~\ref{Fig:squid_graph_approximate} and in the $4^{th}$-order subgraphs, with the exception the subgraphs in the bottom-left corner of Figure~\ref{Fig:squid_graph_approximate}. In a practical device isolation can however be restored for $\gamma_C=\gamma_D$ if the resonant coupling rate $\beta^{cd}_{\omega,\omega}$ is increased. In a real device we expect such a rate to be higher, due to the stray capacitive coupling between the input and output as well as other parasitics \cite{ranzani2013broadband}. Finally, in Figure~\ref{Fig:squid_graph_approximate2} we show the computed isolation and power gain of the SQUID as a function of the bias points for the exact numerical solution and for the lowest-order approximations. We find that graphs up to $4^{th}$ order in $\epsilon$ are enough to accurately describe the isolation properties of the SQUID. The exact power gain was also computed numerically by extracting the impedance matrix from the scattering matrix of the SQUID and then calculating the gain as discussed in~\cite{kamal2012gain}. In principle the power gain may also be computed from the subgraphs, but the number of necessary graphs is large, even at lowest order. In this system, graphs help identify the elements of a complex multimode parametric scattering problem that contribute to the isolation properties. 

\section{Conclusion} \label{section:conclusion}
In this work we introduced a graph theoretical representation that can be used as a combinatorial accounting tool to analyze nonreciprocity in coupled-mode systems and used it to derive a minimal unilateral parametric amplifier and analyze the condictions for nonreciprocity in SQUID amplifiers. An abstract graph associated to the coupled-mode system was used to compute the input/output scattering coefficients between modes and derive general constraints that need to be satisfied in order for nonreciprocity to occur. In order for a multi-mode system to be reciprocal the sum of the phases along any loop has to be $0$ or $\pi$. If this condition is violated, phase and/or amplitude nonreciprocity is present and this condition can be interpreted as interference between different permutations of connecting paths. We also find that dissipation in the remaining disjoint vertices/modes is \emph{crucial} for amplitude nonreciprocity/isolation. Specifically, it breaks the symmetry between forward and backward multi-mode scattering processes such that one may obtain constructive loop interference in one direction, and destructive interference in the reverse direction. As a result we were able to design a novel three-mode parametric amplifier, characterized by forward gain, input match and reverse isolation that can be integrated on a superconducting chip. 

Although the scattering matrix of a multi-mode system can always be computed numerically using more conventional approaches, graphs offer a new perspective that can be useful in the problem of synthesizing new multi-mode systems and devices. For instance, this can be leveraged as a design tool, allowing one to reduce a set of desired device characteristics to a minimal description that can be utilized to design new amplifier/converter concepts such as the Delta amplifier. As an approach, it is conceptually similar to the analysis of electric networks by means of circuit representations of the network components, where graphical manipulation rules are used to simplify the circuit analysis. We also showed, with the complex example of the DC-SQUID, that graphs aid in identifying the critical elements that create isolation in the presence of forward gain. Finally, graphs may provide a useful approach to engineering nonreciprocity in recent multi-mode parametrically coupled hybrid systems such as the electrical-mechanical-optical (three-mode) transduction bridge \cite{andrews2014bidirectional} and electro-mechanical amplifiers \cite{metelmann2014quantum}. In general, problems like these are particularly well-suited to a graph-based approach since many of the interesting properties one is interested in (gain, isolation, gain-bandwidth product, \textit{etc.}) actually arise from the structure of the coupling network (the topology of the graph) and not the physical particulars of a given implementation.

\section*{Acknowledgements}

The authors would like to thank Prof.\,M.\,Devoret (Yale) and Dr.\,Archana Kamal (MIT) for motivating discussions.

\appendix
\section{Derivation of the scattering matrix}

In this section we formally derive Equation~\ref{eq:scattering}. We begin by considering a set of $N_r$ coupled resonators described by:

\begin{equation}
\label{system_eq}
\frac{dA}{dt}=f(A,A^{in}),
\end{equation}

where $A$ is a vector of $N_r$ internal complex normal mode amplitudes and $A^{in}$ describes the set of input signal drives for each of these modes. We assume that $A$ and $A^{in}$ can be expressed as a sum of periodic functions. In this case harmonic balance can be used to rewrite Equation~\ref{system_eq} as a system of $N_m$ coupled-mode equations:

\begin{equation}
\label{eq:coupled_modes}
\frac{dB}{dt}=-iHB + KB^{in},
\end{equation}

where $B^{in}=[{b^{in}_1,b^{in}_p,\dots,b^{in\dagger}_{p+1},\dots,b^{in\dagger}_{N_m}}]$ is a vector of $N_m=p+q$ input stimuli and $B=[{b_1,b_p,\dots,b^{\dagger}_{p+1},\dots,b^{\dagger}_{N_m}}]$ is a vector of $N_m$ internal modes that can be expressed as:

\begin{align}
\label{eq:modes}
b_j &= \tilde b_j e^{-i\omega^s_j t}, \\
\label{eq:modes2} 
b^{in}_j &= \tilde b^{in}_j e^{-i\omega^s_j t} for 1\leq j\leq p, \\
\end{align}

and the corresponding conjugate expressions for $p+1 \leq j \leq N_m$. Here, we allow for the number of modes to exceed the number of resonators, $N_m\ge N_r$ and label the drive/response frequencies with an $s$, while the mode itself is labeled with a subscript $j$ so that ``$\omega^s_j$'' denotes the drive/response frequency of mode $j$ which can be based within any of the $N_r$ resonators depending on how the coupling is directed within the function $f(a,a^{in})$. The diagonal matrix $K=\operatorname{diag}(\sqrt{\gamma^{ext}_1},...,\sqrt{\gamma^{ext}_{N_m}})$ describes coupling to the environment through the external dissipation rates $\{\gamma^{ext}_{j}\}$. The diagonal elements $h_{jj}$ of the $N_m\times N_m$ coupling matrix $H$ can then be expressed in this mode basis as~\cite{Dykman2012fluctuating}:

\begin{eqnarray}
\label{Helements}
h_{jj} = \left\{
\arraycolsep=1.4pt\def\arraystretch{1.6}
\begin{array}{rll}
\omega_j-i\gamma_j/2, & ~~~\mathrm{for}&1\leq j \leq p\\
-\omega_j-i\gamma_j/2, & &p+1 \leq j \leq N_m,
\end{array}
\right.
\end{eqnarray}

and the off-diagonal elements $h_{jk}$ are given by

\begin{equation}
\label{Helements2}
h_{jk} = 
\frac{1}{2}\left\{
\begin{array}{rll}
g_{jk}e^{i(\omega^s_j-\omega^s_k)t} + \mathrm{c.c.} & \mathrm{for}&1 \leq j,k \leq p\\
g_{jk}e^{i(\omega^s_j+\omega^s_k)t} + \mathrm{c.c.} & & 1 \leq j \leq p,\\
	& & ~p+1 \leq k \leq N_m\\
-g_{jk}e^{i(\omega^s_j+\omega^s_k)t} + \mathrm{c.c.}& & 1 \leq k \leq p,\\
 	& & ~p+1 \leq j \leq N_m\\
-g_{jk}e^{i(\omega^s_j-\omega^s_k)t} + \mathrm{c.c.}& &p+1 \leq j,k \leq N_m.\\
\end{array}
\right.
\end{equation}

In Equations~\ref{Helements} and \ref{Helements2} $\omega_j$ and $\gamma_j$ are the natural frequency and total dissipation rate for the resonator in which mode $j$ resides, while the ``*'' indicates complex conjugation, and 
$g_{jk}$ is the complex coupling coefficient between modes $j$ and $k$. We further assume that $g_{jk}=g^*_{kj}$ if $1 \leq j,k \leq p$ or $p+1 \leq j,k \leq N_m$ and $g_{jk}=-g^*_{kj}$ for $1 \leq j \leq p$ and $p+1 \leq k \leq N_m$, or for  $1 \leq k \leq p$ and $p+1 \leq j \leq N_m$. For example, the coupling matrix for a 2-mode frequency converter or resonantly coupled oscillator system corresponds to the case $p=2, q=0$, while a conventional parametric amplifier couples two modes, one at a positive frequency ($p = 1$) and the other at a negative frequency ($q = 1$)~\cite{yurkeSU2}. With this general prescription for defining our mode basis, we can perform an input/output analysis of the system (Equation ~\ref{eq:coupled_modes}) and calculate the vector of output (scattered) fields $b^{out}$ by means of the relation~\cite{Dykman2012fluctuating}:

\begin{equation}
\label{inputoutput}
b_j^{in}-e^{i\xi_j} b_j^{out}= \sqrt{\gamma_j^{ext}}b. 
\end{equation}

In general, the phase factor $\xi_j$ depends on the nature of the coupling to mode $j$. For simplicity, we set $e^{i\xi_j} = -1$, as one would obtain with a small series capacitance or weakly reflecting mirror. Other choices of coupling will have the effect of modifying the identity matrix on the r.h.s.~of Equation~\ref{eq:scattering}. By substituting Equation~\ref{eq:modes} and~\ref{eq:modes2} into Equation~\ref{eq:coupled_modes} and by use of the input/output boundary conditions (Equation~\ref{inputoutput}), we can obtain the expression for the scattering matrix in Equation~\ref{eq:scattering}. 

\section{Coupled-modes analysis of the DC-SQUID}
A DC-SQUID consists of a pair of Josephson junctions inside a superconducting loop as in Figure~\ref{Fig:squid}. Each junction is characterized by a phase difference $\phi_{1,2}$ across its terminals.  Defining the common and differential phase difference $\phi_{C,D}=\phi_1 \pm \phi_2$, the dynamical equations describing the system are~\cite{kamal2012gain}:

\begin{align}
\label{squid_equations}
\frac{d\phi_c}{dt} &= \frac{\omega_B}{2} -\omega_c\sin(\phi_c)\cos(\phi_d) \\
\frac{d\phi_d}{dt} &= \frac{2R}{L}(\pi\phi_e-\phi_d)-\omega_c\cos(\phi_c)\sin(\phi_d),
\end{align} 

where $\omega_B=2\pi I_b R / \Phi_0$, and $\omega_c=2\pi I_c R / \Phi_0$. $I_c$ is the junction critical current and $I_b$ is the bias current of the SQUID, $\Phi_0$ is the magnetic flux quantum, $L$ is the loop geometric inductance and $R$ is the equivalent shunt resistance across the SQUID. The system~(\ref{squid_equations}) can be linearized around the following bias point:

\begin{align}
\label{squid_phases}
\phi_c &= \omega_J t+\delta_c \\
\phi_d &= \phi_e+\delta_d,
\end{align}

where $\delta_c,\delta_d$ are small perturbations. In~\cite{kamal2012gain} it was shown that by applying a harmonic balance procedure, the phases $\delta_{c,d}$ can be expressed as

\begin{equation}
\label{squid_pumps}
\delta_{c,d}=p_{c,d}+ s_{c,d} = \sum \limits_{k=-M}^M {p_k^{c,d}e^{ik\omega_Jt}+s_k^{c,d}e^{i(\omega+k\omega_J)t}},
\end{equation} 

where the components $p_k^{c,d}$ are internally generated pumps at frequency $k\omega_J=2\pi kV_{dc}/\Phi_0$, where $V_{dc}$ is the static DC voltage across the SQUID. The pumps cause parametric frequency mixing between the various signal modes $s_k^{c,d}$. In~\cite{kamal2012gain} the harmonic balance equations were solved for the case $M=3$ to show how nonreciprocity arises from the interference between the different parametric conversion paths. As shown in the main text, we can employ the graph representation of a coupled-mode system to write explicitly the interference conditions needed to maximize the amount of amplitude nonreciprocity between two modes. In the $M=3$ case we can describe the SQUID as a 10-mode system. The 10 modes have frequencies ${\omega \pm 2\omega_J,\omega \pm \omega_J,\omega}$ and are doubly degenerate (for each frequency there are a common and a differential excitation). In order to calculate the scattering parameters, we need to find the coupling coefficients between the modes as a function of the internal pump amplitudes. This can be obtained by linearising Equations~(\ref{squid_equations}) for small signal mode-amplitudes and equating terms at the same frequency in order to obtain a set of coupled mode equations~\ref{eq:coupled_modes}. After taking symmetries into account, the system can be described by the following five coupling coefficients

\begin{eqnarray}
\label{coupling_equations}
g^{cc}_{\omega,\omega-\omega_J}&=&i\epsilon(\cos{\phi_e}-ip_2^{c}\cos{\phi_e}-p_2^{d}\sin{\phi_e}) \\ 
g^{cd}_{\omega,\omega-\omega_J}&=&i\epsilon(i\sin{\phi_e}-p_2^{c}\sin{\phi_e}-ip_2^{d}\cos{\phi}) \\
g^{cc}_{\omega,\omega-2\omega_J}&=&i\epsilon(ip_1^{c}\cos{\phi_e}-p_1^{d}\sin{\phi_e}) \\
g^{cd}_{\omega,\omega-2\omega_J}&=&i\epsilon(ip_1^{d}\cos{\phi_e}-p_1^{c}\sin{\phi_e}) \\
g^{cd}_{\omega,\omega}&=&i\epsilon(\operatorname{Im}(p_1^d)\cos{\phi_e}-\operatorname{Re}(p_1^c)\sin{\phi_e}),
\end{eqnarray} 

where for example $g^{cc}_{\omega,\omega-\omega_J}$ corresponds to the coupling between the common modes at frequency $\omega$ and $\omega-\omega_J$.

\section*{References}

\providecommand{\newblock}{}

\end{document}